\newcommand{\leappr}{\raisebox{-1ex}{    $\stackrel{\textstyle    <}{\sim}$   }}
\newcommand{\geappr}{\raisebox{-1ex}{    $\stackrel{\textstyle    >}{\sim}$   }}
\newcommand{\omi}{\Omega_i}                          

\newcommand{\nuii}{\nu_{ii}}                        
\newcommand{\nuei}{\nu_{ei}}
\newcommand{\nuen}{\nu_{en}}                        
\newcommand{\nuin}{\nu_{in}}
\newcommand{\tez}{T_{e0}}                                
\newcommand{\tiz}{T_{i0}}
\newcommand{\ds}[1]{\displaystyle{#1}} 
\newcommand{\ignore}[1]{}
\renewcommand{\ni}{n_{i1}}
\renewcommand{\ne}{n_{e1}}
\newcommand{\niz}{n_{i0}}
\newcommand{\nez}{n_{e0}}
\newcommand{\ui}{u_{i\parallel1}}
\newcommand{\ue}{u_{e\parallel1}}
\newcommand{\uez}{u_{e\parallel0}}
\newcommand{\pdt}[1]{{\partial #1\over\partial t}}
\newcommand{\del}{\nabla_\parallel}
\newcommand{\ti}{T_{i1}}
\newcommand{\te}{T_{e1}}
\newcommand{\mup}{\mu_\parallel}
\newcommand{\muh}{\hat\mu_\parallel}
\newcommand{\nueiz}{\nu_{ei0}}
\newcommand{\chie}{\chi_{e\parallel}}
\newcommand{\chii}{\chi_{i\parallel}}
\newcommand{\nti}{\tilde n_i}
\newcommand{\nte}{\tilde n_e}
\newcommand{\omz}{\omega_{\parallel0}}
\newcommand{\tti}{\tilde T_i}
\newcommand{\tte}{\tilde T_e}
\newcommand{\phit}{\tilde\varphi}
\newcommand{\chih}{\hat\chi_e}
\newcommand{\chieh}{\hat\chi_{e\parallel}}
\newcommand{\chiih}{\hat\chi_{i\parallel}}
\newcommand{\ih}{\hat I_0}
\newcommand{\kp}{k_\parallel}
\newcommand{\qdz}{Q_{d0}}
\newcommand{\zdz}{Z_{d0}}
\newcommand{\ndz}{n_{d0}}

\newcommand{\bm}[1]{\mbox{\boldmath$#1$}}

\documentstyle[prb,aps,multicol]{revtex}

\begin{document}

\title{Electron-drift   driven  ion-acoustic   mode  in  a  dusty  plasma  with
collisional effects}

\author{R. Singh and M. P. Bora\footnote{Permanently at {\it Physics Department, 
Gauhati University, Guwahati-781~014, India.}\\
Electronic mail~:~{\tt mbora@gw1.dot.net.in}}}
\address{Institute for Plasma Research, Bhatt, Gandhinagar-382~428, India.}

\maketitle

\begin{abstract}
Instabilities of ion-acoustic waves in a dusty plasma with electron-drift, collisional, and
dust charge fluctuations effects, have been investigated. The regimes are clearly marked out
where the theory is applicable. The critical electron-drift velocity required to drive the 
instability is predicted. It is also shown that electron thermal conductivity and charged 
grains concentration enhance the growth of the ion-acoustic mode whereas ion-viscosity,
ion-thermal conductivity, and dust charge fluctuations have a stabilizing effect.

\end{abstract}

\bigskip\noindent


\tighten
\begin{multicols}{2}
\section{Introduction}  Dusty  plasmas  have  acquired  considerable  importance
\cite{shukla1}  because of their  applications  to  astrophysics  and planetary
physics \cite{goertz1}, problems of plasma processing\cite{graves1}, and physics
of strongly coupled systems  \cite{ikezi1}.  Experimentally, one of the important areas  of  investigation  is  the  study  of  low  frequency   ($\omega\leq\omi$)
fluctuations in a plasma with dust grains.  In a recent  experiment by Barkan et
al.  \cite{barkan1},  
it is shown  that the phase
velocity of ion-acoustic wave increases with the density of negatively charged
dust grains and thus ion Landau damping becomes less
severe in a plasma with charged dust grains \cite{barkan1} or with negative ions
\cite{song1}.

The ion-acoustic wave is a typical  compressional mode in which ions provide the
inertia and electrons  bring in the pressure  effects  through  their  shielding
cloud.  In a magnetized  plasma, it propagates  with acoustic  phase speed along
the magnetic field lines.  Earlier, the kinetic theory of the current-driven 
ion-acoustic instability, in a fully ionized, collisional 
\cite{rosenberg1,dangelo1}
and collisionless  \cite{braginskii1}
two-component plasma, has been 
investigated
by various  authors.  In the limit of strong  collisions  (i.e.\  mean free path
smaller than the wavelength),  Coppi et al.\  \cite{coppi1} and Rognlien et al.
\cite{rognlien1}  have used the two-fluid  equations for an  investigation of 
the
current-driven ion-acoustic  instability. Kulsrud and Shen \cite{kulsrud1} have 
investigated the effect of weak collisions on ion-acoustic wave using a 
Fokker-Plank model, while St\'efant \cite{stefant1} used kinetic 
theory and a collision integral of Bhatnagar-Gross-Krook type to show that 
ion-viscosity leads to a damping of the ion-acoustic instability, whereas,
the electron-ion 
collisions tend to increase the growth rate. Recently, Rosenberg  
\cite{rosenberg1}
has carried out the  investigation of  current-driven  ion wave instability in a
fully ionized  collisionless dusty plasma using kinetic theory to
predict the critical electron-drift velocity relative to the ions, required to drive the
instability.

Interpretations  of ion-acoustic  wave  excitation in a dusty plasma  experiment
have  typically  relied on the  theories  based  on collisionless
inverse electron Landau damping  \cite{coppi1,rognlien1}.  A realistic  analysis
of the  experimental  situation shows that in a typical experimental
setup\cite{barkan1,song1}, the plasma density {\it may\/} be quite high
$\sim10^9$-$10^{10}\, \rm cm^{-3}$ when the plasma temperature is quite low.
As an example,
when plasma  density  $\sim10^{10}\,  \rm cm^{-3}$, with  electrons and ions 
having
approximately   equal   temperatures   $T_e\sim   T_i\sim0.1\,   \rm  eV$,  the
electron-ion  collision  mean free path  ($\lambda_e$) may be comparable or even
smaller  than the plasma  scale length  ($L$) along the  magnetic  field,  i.e.\
$\lambda_e\leappr  L$.  Under these  conditions,  the electrons no longer follow
the Boltzmann relation and a collisionless theory of the electron-drift driven 
ion-acoustic  instability is  not justified.  It is therefore  important to 
re-examine
this ion wave instability in a collisional dusty plasma.
 In these experiments cited above, however, ion-acoustic waves are launched by
applying a sinusoidal voltage to a grid immersed in plasma rather than by
drawing an electric current. Nevertheless, the effect of Coulomb collisions
in presence of current driven
ion-acoustic waves in such a plasma, as is considered in this work, remains to
be seen in a laboratory experiment. In another experiment by Barkan et
al.\cite{barkan2}, current driven ion-acoustic waves in a dusty plasma are
studied. But, this experiment was conducted at a relatively low plasma
densities
$\sim10^6$-$10^7\, \rm cm^{-3}$. 

In this paper, we investigate an electron-drift  (or electron current) driven 
ion-acoustic  mode in the  presence of  negatively  charged  dust  grains.  The 
dust
charged  grains  are  considered  to be massive  particles  in a  multicomponent
plasma.  This is valid when $a\ll  d\ll\lambda_D$,  where $a$ is the average  
dust
radius, $d$ is the  average distance between the dust particles,
and  $\lambda_D$ is the 
plasma
Debye length.

The paper is organized as follows. In Sec.II, the regimes of interest, where the
theory is applicable are discussed. Sec.III deals with the linear instability
theory. Sec.IV contains the conclusions.


\section{Regimes  of  interest}
At first, we refer to regimes of interest  where
the  present  calculations  are  relevant. 
We consider a typical experimental situation such as considered in experiments
by Barkan et al.\cite{barkan1,barkan2} and Song et al.\cite{song1}. In such  conditions,  the
typical parameters in these experiments are~:  potassium plasma ($\rm K^+$) with
$m_i/m_p\sim40$,  $m_i$ and $m_p$  being the ion and proton  masses  
respectively,
dust mass $m_d\sim10^{-12}\, \rm gm$, gas pressure  $\leappr10^{-5}\, \rm Torr$,
the corresponding neutral ($n$) density $n_n\simeq2\times10^{11}\, \rm cm^{-3}$,
equal electron ($e$) and ion ($i$) temperatures  $T_e\sim  T_i\sim0.1\, \rm eV$,
and  dust  grain  temperature  $T_d\sim0.03\,  \rm  eV$  (assumed  to be at room
temperature).  The plasma is confined  radially by a magnetic field  $B\sim0.4\,
\rm T$ and the plasma length along the magnetic  field  $L\sim40$~cm  with a
radius   $r_p\sim2\,  \rm  cm$  for  the  plasma  column.  The   characteristic
frequency  of  the  waves   $f\sim$~20-80~kHz   which corresponds  to  
$\omega=2\pi
f\sim$~1-5$\times10^5\,  \rm rad/s$.  The other plasma parameters are~: 
electron-thermal   velocity   $c_e=(T_e/m_e)^{1/2}\sim   1.3\times10^7\,  \rm  cm/s$,
ion-thermal  velocity   $c_i=(T_i/m_i)^{1/2}\sim5\times10^4\,   \rm  cm/s$,  
the ion-acoustic  speed  $c_s=(T_e/m_i)^{1/2}\sim5\times10^4\,  \rm cm/s$, the dust-thermal
velocity   $c_d=    (T_d/m_d)^{1/2}\sim2\,    \rm   cm/s$,   the ion   gyrofrequency
$\omi=eB/m_ic\sim10^6\,         \rm        s^{-1}$,         the ion Larmor        radius
$a_i=c_i/\omi\sim7\times10^{-2}\,   \rm   cm$,   the collision   frequency
of electrons and ions with the neutrals $\nuen=n_n\sigma_{en}c_e\sim1.3\times10^4\,  
\rm s^{-1}$, 
$\nuin=n_n\sigma_{in}c_i\sim50\,      \rm     s^{-1}$      (where
$\sigma_{in}\sim\sigma_{en}\sim5\times10^{-15}\,  \rm cm^2$),  with the corresponding
mean free  paths  $\lambda_{en}\sim\lambda_{in}\sim10^3\,  \rm cm$.  The  
$e$-$i$
collision  frequency  ($\nuei$) and $i$-$i$ collision frequency ($\nuii$) in the
form presented by Braginskii  \cite{braginskii1}  are calculated  for the 
present
parameters and are found to be as $\nuei=1/\tau_e\sim8.4\times10^6\, \rm s^{-1}$ and $\nuii=
1/\tau_i\sim2.8\times10^4\, \rm s^{-1}$ and the mean free paths are  $\lambda_e=c_e/\nuei
\sim1\,    \rm   cm$   and    $\lambda_i=c_i/\nuii\sim1\,    \rm   cm$,   where
$\tau_e=3.5\times10^4\tez^{3/2}/n\lambda$,                  
$\tau_i=3.0\times10^6
(m_i/2m_p)^{1/2}\tiz^{3/2}/n\lambda$,                   
and
$\lambda=23.4-1.15\log n+3.45\log T\sim9.5$ (for $T<50\, \rm eV$) is the Coulomb 
logarithm.

From these calculations the following assumptions can now be made
\begin{description}\parskip -6pt

\item[(1)] the collision mean free paths of electrons and ions are comparable to or even 
smaller than the plasma scale length along the magnetic field i.e.\ 
$\lambda_{e,i}<L$, so that we can make use of Braginskii's fluid equations,

\item[(2)] since $\omega<\nuei$, we can neglect the electron inertia,

\item[(3)] since $a_i<r_p$ and assuming a homogeneous plasma, the ion motion transverse to 
the magnetic field can be neglected (the polarization drift),

\item[(4)] since $\nuei\gg\nuen, \nu_{ed}$ and $\nuii\gg\nuin, \nu_{id}$, so the momentum 
and energy losses of electrons and ions to the neutrals and dust grains can 
also be neglected in the fluid equations,

\item[(5)] since the phase velocity of the observed ion-acoustic wave is much larger than 
the dust-thermal speed i.e.\ $v_p\gg c_d$, we can neglect the dust 
dynamics in the analysis.
However, we shall incorporate the self-consistent fluctuation of charge 
on dust grains, which is shown to cause a damping effect on 
the ion-acoustic wave \cite{Li1,dangelo2}. The charge fluctuation on the 
surface of the dust grains, which
depends crucially on the dust
size and the plasma density, is important when charging rate is comparable 
to the growth of the mode.

\item[(6)] we consider negatively charged grains as an additional charged plasma 
species of uniform massive particles similar to a multicomponent plasma. This 
remains valid as long as $a\ll d\ll\lambda_D$.

\end{description}

\section{Instability analysis}
We write down the basic equations in order to derive the linear dispersion 
relation of electron-drift driven ion-acoustic wave instability including the 
effects of dust charge fluctuations, electron and ion temperature  perturbations, 
ion-viscosity etc. We consider a one-dimensional, plane, homogeneous plasma with 
low $\beta$ ($\beta=4\pi nT_e/B^2\ll1$ is the ratio of kinetic energy to 
magnetic energy). The resultant governing equations for electrons and ions are 
the Braginskii's fluid equations viz., equations of continuity, momentum, and 
energy.

We write down the equilibrium equations that define this specific regime of
interest and confine ourselves to the parallel motions only. They are the ion and
electron continuity equations
\begin{eqnarray}
\frac{\partial n_{i}}{\partial t}+\nabla _{\parallel }\cdot (n_{i}\bm {u}_{i\parallel }) & = & 0,\\
\frac{\partial n_{e}}{\partial t}+\nabla _{\parallel }\cdot (n_{e}\bm {u}_{e\parallel }) & = & 0,
\end{eqnarray}
parallel electron and ion momentum equations
\begin{eqnarray}
m_{i}n_{i}\frac{d\bm {u}_{i\parallel }}{dt} & = & -\nabla _{\parallel }p_{i}+en_{i}\bm {E}+\mu _{\parallel }\nabla ^{2}_{\parallel }\bm {u}_{i\parallel },\\
m_{e}n_{e}\frac{d\bm {u}_{e\parallel }}{dt} & = & -\nabla _{\parallel }p_{e}-en_{e}\bm {E}+\bm {R},
\end{eqnarray}
where $\bm{R}$ represents the momentum gained by the electrons through collision
with the ions\cite{braginskii1},
\begin{eqnarray}
\bm {R} & = & \bm {R}_{u}+\bm {R}_{T},\\
\bm {R}_{u} & \simeq  & -m_{e}n_{e}\nu _{ei}(0.51\bm {u}_{\parallel }),\\
\bm {R}_{T} & \simeq  & -0.71n_{e}\nabla _{\parallel }T_{e},\\
\bm {u}_{\parallel } & = & \bm {u}_{e\parallel }-\bm {u}_{i\parallel }.
\end{eqnarray}
In the above equations, $\bm{E}$ is the electric field driving the instability and
$\mup=0.96\niz\tiz/\nuii$ is the parallel ion-viscosity 
coefficient\cite{braginskii1}. 
In the ion momentum equation, however, we have neglected the collision 
term\cite{coppi1} in the
limit \(\bm {u}_{i\parallel}>(m_{e}/m_{i})(\nu _{ei}/\omega)\bm {u}_{\parallel}\).
We have further neglected the electron viscosity term in Eq.(4). The remaining
equations are the energy equations\cite{braginskii1},
\begin{eqnarray}
\frac{3}{2}n_{i}\frac{dT_{i}}{dt}+p_{i}(\nabla _{\parallel }\cdot \bm {u}_{i\parallel }) & = & \nabla _{\parallel }\cdot (\chi _{\parallel }^{i}\nabla _{\parallel }T_{i}),\\
\frac{3}{2}n_{e}\frac{dT_{e}}{dt}+p_{e}(\nabla _{\parallel }\cdot \bm
{u}_{e\parallel  }) & = & \nabla _{\parallel }\cdot (\chi _{\parallel
}^{e}\nabla _{\parallel }T_{e})\nonumber\\
&& -\,0.71\nabla _{\parallel }\cdot (n_{e}T_{e}\bm {u}_{\parallel }),
\end{eqnarray}
where $\chii=3.9\niz\tiz/m_i\nuii$ and $\chie=3.2\nez\tez/m_e\nuei$ are the 
parallel ion and electron-thermal conductivities, respectively.  In the above
equations, the terms $\sim\omega_{ce,i}^{-1}$ have neglecetd owing to the fact
that $\omega<\omega_{ce,i}$.  

We consider a small electrostatic perturbation and linearize the above equations.
The linearized equations are,
\begin{eqnarray}
\pdt{\ni}+\del(\niz\ui) &=& 0,
\label{i_cont}
\\
\pdt{\ne}+\del(\nez\ue+\ne\uez) &=& 0,
\label{e_cont}
\end{eqnarray}
where equilibrium and perturbed quantities are defined by the subscripts 0 and 
1, respectively and $\uez$ is the zeroth order electron-drift velocity with 
respect to the ions and dust grains.
The electrons drift with respect to the ions and dust particles, so that 
\( \bm {u}_{i0}=\bm {u}_{d0}=0 \).
The ion motion along the magnetic field $B$ 
is given by the linearized parallel momentum equation
\begin{eqnarray}
m_i\niz\pdt{\ui} &=& -\del(e\niz\varphi_1+\niz\ti+\ni\tiz)\nonumber\\
&& +\,\mup\del^2\ui,
\label{i_mom}
\end{eqnarray}
where 
$\varphi_1$ is the perturbed electrostatic potential. For $\omega<\nuei$, the 
parallel 
electron momentum equation becomes
\begin{eqnarray}
0 &=& \del(e\nez\varphi_1-\ne\tez-\nez\te)-0.71\nez\del\te
\nonumber
\\
&& -\, 0.51m_e\nueiz(\nez\ue-\nez\ui)\nonumber\\
&& -\,0.51m_e\nez\uez\nu_{e1},
\label{e_mom}
\end{eqnarray}
where 
$\nu_{e1}=\nueiz(\ne/\nez-3\te/2\tez)$ 
and 
the equilibrium drift velocity $\uez=-eE_{\parallel0}/m_e\nuei$. The perturbed 
electron and ion temperature equations are given by
\end{multicols}
\vskip-10pt
\vrule depth 10pt width 0pt\hfill\vrule height 10pt depth 0pt
\vskip-10pt\hrule
\begin{eqnarray}
{3\over2}\niz\pdt{\ti}+\niz\tiz\del\ui &=& \chii\del^2\ti,
\label{i_energy}
\\
{3\over2}\nez\left(\pdt{}+\uez\del\right)\te+\nez\tez\del\ue &=& \chie\del^2\te 
-0.71\nez\tez\del(\ue-\ui)
\nonumber
\\
&& -\, 0.71\uez\del(\nez\te+\ne\tez),
\label{e_energy}
\end{eqnarray}
\vskip-10pt
\noindent\vrule depth 10pt height 0pt\hfill\vrule height10pt depth0pt width0pt
\vskip-10pt\hrule
\begin{multicols}{2}
\noindent
Finally, we 
write the  equations for dust charge fluctuations as given by Jana et 
al.\cite{jana1,vinod1} and the quasineutrality condition as
\begin{eqnarray}
\left(\pdt{}+\eta\right)Q_{d1} &=& 
-|I_e|\left({\ni\over\niz}-{\ne\over\nez}\right),
\label{fluct}
\\
e(\ne-\ni)+n_{d0}Q_{d1} &=& 0,
\label{quasi}
\end{eqnarray}
where $\qdz=e\zdz$, $\zdz$ is the equilibrium charge number on the surface of 
the dust grains, $Q_{d1}=eZ_{d1}$, $Z_{d1}$ is the charge fluctuation, 
$\eta=e|I_e|(\tez^{-1}-W_0^{-1})/C$ with \mbox{$W_0=\tiz-e\phi_{f0}$}, $\phi_{f0}$
is
the potential at the dust surface, $|I_e|\sim I_i\sim e\niz\pi a^2c_s$ is the 
equilibrium electron (or ion) current at the dust surface, and $C\sim a$ being the 
grain capacitance. In writing Eq.(\ref{quasi}), we have neglected the dust 
density fluctuations as the ion transit time ($\sim L/c_s$) is much shorter than 
the dust transit time \cite{kaw2} [$\sim L/c_{d\ast}$, 
$c_{d\ast}\sim(\zdz^2\ndz\tiz/\niz m_d)^{1/2}$] along the magnetic field.

It should, however, be noted that in writing the above equations, we have 
assumed a stationary background equilibrium. On the other hand, a non-stationary 
background equilibrium may lead to stabilization of the ion-acoustic instability 
in a low temperature collisional plasma \cite{kaw1}.

We now take perturbations of the form
\begin{equation}
f(z,t)\sim f_1e^{-i(\omega t-\kp z)}
\label{perturb}
\end{equation}
and write Eqs.(\ref{i_cont}-\ref{quasi}) as
\end{multicols}
\vskip-10pt
\vrule depth 10pt width 0pt\hfill\vrule height 10pt depth 0pt
\vskip-10pt\hrule
\begin{eqnarray}
\nti &=& {\kp\ui\over\omega},
\label{ni_pert}
\\
\nte &=& {\kp\ue\over(\omega-\omz)},
\label{ne_pert}
\\
\kp\ui &=& {\kp^2c_s^2\over(\omega+i\muh)}\left(\phit+{\nti\over\tau} 
+{\tti\over\tau}\right),
\label{ui_pert}
\\
\nte+\left(1.71+i{3\over2}{\omz\over\chih}\right)\tte-\phit &=& 
i{\kp\over\chih}(\ue+\uez\nte-\ui),
\label{ue_pert}
\\
\left({3\over2}\omega+i\chiih\right)\tti &=& \kp\ui,
\label{ti_pert}
\\
\left({3\over2}\omega-2.21\omz+i\chieh\right)\tte &=& \kp(1.71\ue-0.71\ui) 
+0.71\omz\nte,
\label{te_pert}
\end{eqnarray}
where $\tau=\tez/\tiz$, $\nte=\ne/\nez$, $\nti=\ni/\niz$, $\tti=\ti/\tiz$, $\tte=\te/\tez$, 
$\phit=e\varphi_1/\tez$ are the normalized
 variables with
$\chih\simeq\kp^2c_e^2/0.51\nuei$, $\muh=0.96\kp^2c_i^2/\nuii$, $\omz=\kp\uez$, 
$\chieh=3.2\kp^2c_e^2/\nuei$, and $\chiih=3.9\kp^2c_i^2/\nuii$.
We can now combine Eqs.(\ref{fluct}) and (\ref{quasi}) as
\begin{equation}
\left(1+i{\zdz\ndz\over\nez}{\ih\over(\omega+i\eta)}\right)\nte 
=\left({\niz\over\nez}+i{\zdz\ndz\over\nez}{\ih\over(\omega+i\eta)}\right)\nti,
\label{quasi2}
\end{equation}
where $\ih=|I_e|/e\zdz$ and $\lambda_D=(\tez/4\pi\nez e^2)^{1/2}$ is the plasma 
Debye length.

In the limit
$\omega\sim\omz<\chieh$  i.e.\ 
$3.2(\kp\lambda_e)^2\nuei/\omega>1$, 
Eq.(\ref{te_pert}) can be re-written, using Eq.(\ref{ne_pert}), as
\begin{equation}
\tte=-i{(1.71\omega-\omz)\over\chieh}\nte+i0.71{\omega\over\chieh}\nti.
\label{te_pert2}
\end{equation}
From Eqs.(\ref{ni_pert}) and (\ref{ti_pert}), we get
\begin{equation}
\tti={\omega\over\left({3\over2}\omega+i\chiih\right)}\nti.
\label{ti_pert2}
\end{equation}
Substituting Eqs.(\ref{ui_pert}) and (\ref{ti_pert2}) into Eq.(\ref{ni_pert}), 
we obtain
\begin{equation}
\left[1-{\kp^2c_i^2\over\omega(\omega+i\muh)}-{\kp^2c_i^2\over 
(\omega+i\muh)\left( {3\over2}\omega+i\chiih\right)}\right]\nti 
={\kp^2c_s^2\over\omega(\omega+i\muh)}\phit.
\label{ni_phi}
\end{equation}
Using Eqs.(\ref{ni_pert}), (\ref{ne_pert}), and (\ref{te_pert2}), 
Eq.(\ref{ue_pert}) can be written as
\begin{equation}
\left[1-i{\omega\over\chih}-i1.71{(1.71\omega-\omz)\over\chieh}\right]\nte 
-\phit=-i1.75{\omega\over\chih}\nti.
\label{ne_phi_ni}
\end{equation}
In the limit of $\omega>\eta$, Eqs.(\ref{quasi2}), (\ref{ni_phi}), and 
(\ref{ne_phi_ni}) can be combined to obtain the following dispersion relation,
\begin{eqnarray}
\left[1-{\kp^2c_i^2\over\omega(\omega+i\muh)}-{\kp^2c_i^2\over 
(\omega+i\muh)\left( {3\over2}\omega+i\chiih\right)}\right] 
\left[1+i{\zdz\ndz\over\nez}{\ih\over\omega}\left(1-{\nez\over\niz}\right) 
\right]
\nonumber
\\
={\niz\over\nez}{\kp^2c_s^2\over\omega(\omega+i\muh)} 
\left[1-i{\omega\over\chih}-i1.71{(1.71\omega-\omz)\over\chieh}\right.
\nonumber
\\ +\, \left.i1.75{\omega\over\chih}{\nez\over\niz}\left\{1+i{\zdz\ndz\over\nez} 
{\ih\over\omega}\left(1-{\nez\over\niz}\right)\right\}\right].
\label{dispersion}
\end{eqnarray}

For $\tiz\sim\tez$ and $\kp^2c_e^2/\omega\nuei>1$, it is obvious that 
$\omega>\muh, \chiih$ (i.e.\ $\kp^2c_i^2/\omega\nuii<1$). We now write 
Eq.(\ref{dispersion}) as $\epsilon(\kp,\omega)=0$ and $\omega=\omega_r+i\gamma$ 
with $\omega_r>\gamma$. Taking $\varepsilon=\muh/\omega_r^{(0)}$ as an expansion
parameter, the real part of $\omega$, to the first order, can be written as,
\begin{equation}
\omega_r^2\simeq(\omega_r^{(0)})^2+1.71\kp^2c_s^2\left({\muh\over\omega_r^{(0)}}
{\niz\over\nez}{\omz\over\chieh}\right),
\label{real}
\end{equation}
where $\omega_r^{(0)}$ is given by
\begin{equation}
\omega_r^{(0)}=\kp c_s\left({\niz\over\nez}+{5\over3}{\tiz\over\tez} 
\right)^{1/2}.
\label{real0}
\end{equation}
The growth rate can now be written as
\begin{eqnarray}
\gamma &\simeq& -\, {\muh\over2}-{2\over9}{\kp^2c_i^2\over\omega_r^2}\chiih 
-{Z_{d0}n_{d0}\over2\nez}{\niz\over\nez}{\kp^2c_s^2\over\omega_r^2}\ih 
\left(1-{\nez\over\niz}\right)
\nonumber
\\
&& +\, 1.71{\niz\over\nez}{\kp^2c_s^2\over2\chieh}\left({\omz\over\omega_r} 
-1-{Z_{d0}n_{d0}\over\niz}{\chieh\over\chih}\right).
\label{imag}
\end{eqnarray}

\begin{multicols}{2}

\begin{figure}
{\small
\begingroup%
  \makeatletter%
  \newcommand{\GNUPLOTspecial}{%
    \@sanitize\catcode`\%=14\relax\special}%
  \setlength{\unitlength}{0.1bp}%
\begin{picture}(2160,1511)(0,0)%
\special{psfile=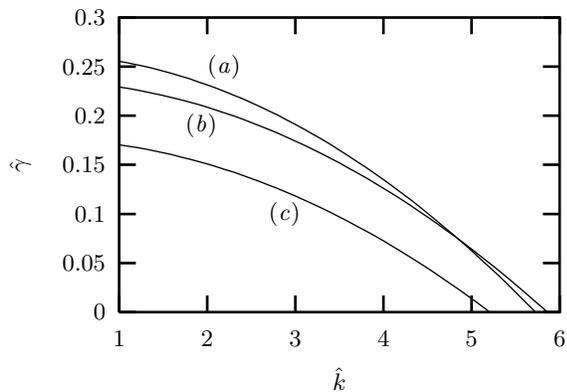 llx=0 lly=0 urx=432 ury=353 rwi=4320}
\put(1014,670){\makebox(0,0)[l]{({\it c\/})}}%
\put(699,1004){\makebox(0,0)[l]{({\it b\/})}}%
\put(782,1226){\makebox(0,0)[l]{({\it a\/})}}%
\put(1280,50){\makebox(0,0){$\hat k$}}%
\put(100,855){%
\special{ps: gsave currentpoint currentpoint translate
270 rotate neg exch neg exch translate}%
\makebox(0,0)[b]{\shortstack{$\hat\gamma$}}%
\special{ps: currentpoint grestore moveto}%
}%
\put(2110,200){\makebox(0,0){6}}%
\put(1778,200){\makebox(0,0){5}}%
\put(1446,200){\makebox(0,0){4}}%
\put(1114,200){\makebox(0,0){3}}%
\put(782,200){\makebox(0,0){2}}%
\put(450,200){\makebox(0,0){1}}%
\put(400,1411){\makebox(0,0)[r]{0.3}}%
\put(400,1226){\makebox(0,0)[r]{0.25}}%
\put(400,1041){\makebox(0,0)[r]{0.2}}%
\put(400,856){\makebox(0,0)[r]{0.15}}%
\put(400,670){\makebox(0,0)[r]{0.1}}%
\put(400,485){\makebox(0,0)[r]{0.05}}%
\put(400,300){\makebox(0,0)[r]{0}}%
\end{picture}%
\endgroup
 }
\caption{Normalized growth rate $\hat\gamma$ of the ion-acoustic wave vs.\
normalized wave number $\hat k$ for different values of the parameter 
$\delta=\nez/\niz$. The curves labeled as ({\it a\/}), ({\it b\/}),
and ({\it c\/}) are for
$\delta=1$ (no dust), 0.4, and 0.3, respectively. The other parameters
are $\hat u=6.0$, $\hat I=0$, and $\hat\lambda_{e,i}=0.01$.}
\end{figure}

\begin{figure}
{\small
\begingroup%
  \makeatletter%
  \newcommand{\GNUPLOTspecial}{%
    \@sanitize\catcode`\%=14\relax\special}%
  \setlength{\unitlength}{0.1bp}%
\begin{picture}(2160,1511)(0,0)%
\special{psfile=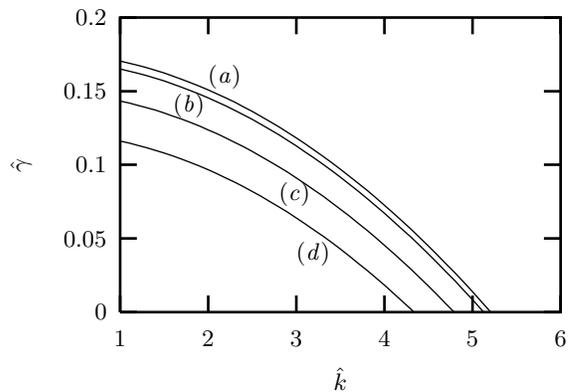 llx=0 lly=0 urx=432 ury=353 rwi=4320}
\put(1114,522){\makebox(0,0)[l]{({\it d\/})}}%
\put(1048,744){\makebox(0,0)[l]{({\it c\/})}}%
\put(649,1078){\makebox(0,0)[l]{({\it b\/})}}%
\put(782,1189){\makebox(0,0)[l]{({\it a\/})}}%
\put(1280,50){\makebox(0,0){$\hat k$}}%
\put(100,855){%
\special{ps: gsave currentpoint currentpoint translate
270 rotate neg exch neg exch translate}%
\makebox(0,0)[b]{\shortstack{$\hat\gamma$}}%
\special{ps: currentpoint grestore moveto}%
}%
\put(2110,200){\makebox(0,0){6}}%
\put(1778,200){\makebox(0,0){5}}%
\put(1446,200){\makebox(0,0){4}}%
\put(1114,200){\makebox(0,0){3}}%
\put(782,200){\makebox(0,0){2}}%
\put(450,200){\makebox(0,0){1}}%
\put(400,1411){\makebox(0,0)[r]{0.2}}%
\put(400,1133){\makebox(0,0)[r]{0.15}}%
\put(400,856){\makebox(0,0)[r]{0.1}}%
\put(400,578){\makebox(0,0)[r]{0.05}}%
\put(400,300){\makebox(0,0)[r]{0}}%
\end{picture}%
\endgroup
 }
\caption{Behavior of $\hat\gamma$ vs.\ $\hat k$ for ({\it a\/})~$\hat I=0$,
({\it b\/})~0.01, ({\it c\/})~0.05, and ({\it d\/})~0.1. Dust density 
parameter $\delta=0.3$ and $\hat\lambda_{e,i}=0.01$.}
\end{figure}

\end{multicols}
\begin{multicols}{2}
We note that, at the zeroth order,
the phase velocity of the ion-acoustic wave is modified with a 
factor of $\left(\ds{\niz\over\nez}+\ds{5\over3}\ds{\tiz\over\tez}\right)^{1/2}$.
The presence of dust (decreasing ratio of $\nez/\niz$) 
leads to an increase in the phase velocity of the ion-acoustic wave.

In the expression for growth rate of the ion-acoustic wave Eq.(\ref{imag}),
the collisional 
damping is represented by the first two factors, the parallel ion-viscosity and 
parallel
 ion-thermal conductivity. The third term causes damping due 
to charge fluctuation on the surface of the dust grains, which is proportional 
to the electron/ion current ($\ih$) on the dust surface. We note that in absence 
of dust charge fluctuation (i.e.\ $\ih=0$), increasing negatively charged dust 
density with respect to electron and ion density, results in an increase of the
growth 
rate of the ion-acoustic instability. This can be seen from Fig.1, where we have 
plotted the normalized growth rate $\hat\gamma=\gamma L/c_s$ against normalized 
wave number $\hat k=k_\parallel L$, as obtained from the solution of the full 
dispersion relation Eq.(\ref{dispersion}), at different negatively charged dust 
concentration $\delta=\nez/\niz$. Note that $\delta=0.3$ means 70\% of the 
electronic charges is now carried by the massive dust particles and $\delta=1$ 
means no negatively charged dust grains. The other parameters in Fig.1 are
$\hat u=\uez/c_s=6$, $\hat\lambda=\lambda_{i,e}/L=0.01$, and $\hat I=\ih L/c_s
=0$. As in ion-acoustic instability without 
dust particles, the positive growth rate, here also, appears only when the 
drifting electron velocity $\uez$, exceeds several times the phase velocity.
Therefore, the necessary condition for required electron-drift relative to ions 
and negatively charged dust grains to excite the ion wave instability is
\end{multicols}
\vskip-10pt
\vrule depth 10pt width 0pt\hfill\vrule height 10pt depth 0pt
\vskip-10pt\hrule
\begin{eqnarray}
\uez &>& c_s\left({\niz\over\nez}+{5\over3}{\tiz\over\tez}\right)^{1/2} 
\left\{\left(1+{Z_{d0}n_{d0}\over\niz}\right)\right.
\nonumber
\\
&& +\, (\kp\lambda_e)^2{\nez\over\niz}{\tiz\over\tez}{\nuei\over\nuii} 
\left[1.8+\left({5\over3}+{\niz\tez\over\nez\tiz}\right)^{-1}\right]
\nonumber
\\
&& +\, \left.1.87\left({Z_{d0}n_{d0}\over\niz}\right)^2{\niz\over\nez} 
\left({\niz\over\nez}+{5\over3}{\tiz\over\tez}\right)^{-1}\left({\pi 
a^2\lambda_e\niz\over Z_{d0}}\right)\left({m_i\over m_e}\right)^{1/2}\right\},
\label{condition}
\end{eqnarray}
\vskip-10pt
\noindent\vrule depth 10pt height 0pt\hfill\vrule height10pt depth0pt width0pt
\vskip-10pt\hrule
\begin{multicols}{2}
\noindent
where we have used, only the zeroth order expression for $\omega_r$.
Note that the ion-viscosity and ion-thermal conductivity 
raise the critical electron current for the ion-acoustic wave to be 
unstable [the term in the square bracket in Eq.(\ref{condition})],
whereas the electron-thermal conductivity tends to increase the growth
rate, as seen from Eq.(\ref{imag}).

The effect of increasing negatively charged dust density on growth rate of the 
ion-acoustic instability may be compensated by the presence of charge 
fluctuation on dust surface, which has a damping effect, as 
shown in Fig.2. In Fig.2 we have taken the dust density parameter $\delta=0.3$. 
The other parameters are same as in Fig.1. We show in Fig.3, the behavior 
of $\hat\gamma$ versus $\delta$ 
for different values of dust current $\hat I=0$, 0.05, 0.1, and 1.0. Other parameters
chosen are $\hat u=6$, $\hat\lambda=0.01$, and $\hat k=5.0$.
The overall behavior is consistent 
with Fig.1 and Fig.2.
\begin{figure}
{\small
\begingroup%
  \makeatletter%
  \newcommand{\GNUPLOTspecial}{%
    \@sanitize\catcode`\%=14\relax\special}%
  \setlength{\unitlength}{0.1bp}%
\begin{picture}(2160,1511)(0,0)%
\special{psfile=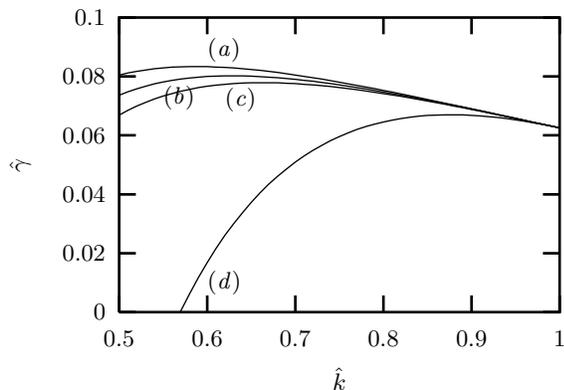 llx=0 lly=0 urx=432 ury=353 rwi=4320}
\put(782,411){\makebox(0,0)[l]{({\it d\/})}}%
\put(848,1100){\makebox(0,0)[l]{({\it c\/})}}%
\put(616,1111){\makebox(0,0)[l]{({\it b\/})}}%
\put(782,1289){\makebox(0,0)[l]{({\it a\/})}}%
\put(1280,50){\makebox(0,0){$\hat k$}}%
\put(100,855){%
\special{ps: gsave currentpoint currentpoint translate
270 rotate neg exch neg exch translate}%
\makebox(0,0)[b]{\shortstack{$\hat\gamma$}}%
\special{ps: currentpoint grestore moveto}%
}%
\put(2110,200){\makebox(0,0){1}}%
\put(1778,200){\makebox(0,0){0.9}}%
\put(1446,200){\makebox(0,0){0.8}}%
\put(1114,200){\makebox(0,0){0.7}}%
\put(782,200){\makebox(0,0){0.6}}%
\put(450,200){\makebox(0,0){0.5}}%
\put(400,1411){\makebox(0,0)[r]{0.1}}%
\put(400,1189){\makebox(0,0)[r]{0.08}}%
\put(400,967){\makebox(0,0)[r]{0.06}}%
\put(400,744){\makebox(0,0)[r]{0.04}}%
\put(400,522){\makebox(0,0)[r]{0.02}}%
\put(400,300){\makebox(0,0)[r]{0}}%
\end{picture}%
\endgroup
 }
\caption{Dependence of $\hat\gamma$ on the dust density parameter $\delta$
for different fluctuation level of the dust charge. ({\it a\/})~$\hat I=0$
i.e.\ no dust charge fluctuation, 
({\it b\/})~$\hat I=0.05$, ({\it c\/})~$\hat I=0.1$, 
and ({\it d\/})~$\hat I=1.0$. Note the change in
behavior of $\hat\gamma$ as the dust current increases. The normalized
wave number $\hat k=5.0$, $\hat u=6.0$, and $\hat\lambda_{e,i}=0.01$.}
\end{figure}
\noindent
However, it is important to note that, although a 
decreasing $\delta$ causes peaking up of the growth rate, at any given wave 
number, a rapid dust charge fluctuation rate (larger $\ih$)
 causes the wave to 
damp at higher dust concentration, which is opposite in behavior to the case of 
absence of dust charge fluctuation (Fig.1 and 2), as can be seen from all the
curves in Fig.3.


For typical experimental parameters $\nez/\niz\sim0.3$, 
$Z_{d0}n_{d0}/\niz\sim0.7$, $\tez\sim\tiz$, and $\nuei/\nuii\sim 
(m_i/m_e)^{1/2}$, 
the necessary threshold condition for instability reduces to

\begin{eqnarray}
\uez &>& 2.24c_s
\!\!\left[1.7+\left({m_i\over m_e}\right)^{1/2}\!\!\!\left(0.7\kp^2\lambda_e^2 
\right.\right.\nonumber\\
&& +\, \left.\left.0.6{\pi a^2\lambda_e\niz\over Z_{d0}}\!\right)\!\right].
\label{threshold}
\end{eqnarray}
It is to be noted that the effects of charge fluctuation at the dust surface 
will tend to play an important role when the mean free paths are comparable to 
the plasma length along the magnetic field and the term
\begin{equation}
{\pi a^2\lambda_e\niz\over\zdz}\geappr3\left({m_e\over m_i}\right)^{1/2}.
\end{equation}
Also notice that in high density and low temperature plasma limits, the 
effects ion-viscosity and thermal conductivity will be important, typically at 
wavelengths such that
\begin{equation}
k_\parallel\lambda_e\sim1.5\left({m_e\over m_i}\right)^{1/4}.
\end{equation}

The condition (\ref{threshold}) can be realized in typical laboratory 
situations\cite{barkan1}. For example, with an average size of dust grains to be
few microns ($a\sim10^{-4}\, \rm cm$) and a corresponding $Z_{d0}\sim10^4$, the condition
(\ref{threshold}) yields
\begin{equation}
\uez\geappr4c_s,
\label{threshold_no}
\end{equation}
for the excitation of the ion-acoustic instability, 
where we have taken the collisional parameters as $\lambda_{e,i}/L\simeq0.01$ and 
$\hat k=k_\parallel L\sim2$. The typical parallel electric field required for
the corresponding threshold electron current is $\sim0.1\, \rm V/m$.

\section{Conclusions}
We have studied the ion-acoustic instability in a collisional dusty plasma with
fluid equations, where
dust grains are treated as massive and  negatively charged component in a multicomponent 
plasma. The regimes are clearly marked out where the theory is applicable,
especially in a relatively high density ($10^{10}\, \rm cm^{-3}$) and low 
temperature plasma where the effect of collisions cannot be neglected.
While treating the negatively charged dust particles, we take into account
the effect of charge fluctuation on the dust surface. We have shown that in such a 
plasma, which are routinely produced in laboratory, there is a significant
impact of electron-ion collisions, even at large mean free paths 
(i.e.\ $\lambda_f\geappr L$). 
We have derived the threshold electron-drift velocity required to drive the ion-acoustic
instability. It is shown that the electron-thermal conductivity and the dust charge 
concentrations reduce the threshold value of electron current for driving the ion-acoustic
mode. In particular, the ion-viscosity and ion-thermal conductivity raise
the threshold current. And a similar effect due to dust charge fluctuations
is also found when the mean free paths are of the order of the plasma length.

\section*{Acknowledgment}
It is pleasure to thank A. Sen for fruitful discussions. One of the authors,
MPB, would also like to thank A. Sen for kind hospitality at IPR.
\end{multicols}


\end{document}